\documentclass[12pt]{JHEP3}
\usepackage[dvips]{epsfig}
\usepackage{epsfig}
\usepackage{graphicx}

\title{Spacetime decay of cones at strong coupling}
\author{Ruth Gregory\\Centre for Particle Theory, University of Durham,
\\South Road, Durham, DH1 3LE, U.K.}
\author{Jeffrey A. Harvey\\Enrico Fermi Institute and Dept. of Physics,
University of Chicago,\\5640 S. Ellis Ave. Chicago, IL 60637, U.S.A.}

\abstract{
We study time dependent solutions in dilaton gravity which correspond to
the decay of conical spacetimes. In string theory this can be interpreted as
a strong coupling limit of the decay 
of a non-supersymmetric orbifold spacetime with localized tachyons. 
}
\preprint{hep-th/0306146\\DCPT-03/66\\IPPP-03/33\\EFI-03-28}

\def\cee{{\relax\hbox{$\inbar\kern-.3em{\rm C}$}}}

\def\ie{{\it i.e.,}}

\newcommand{\be}{\begin{equation}}
\newcommand{\ee}{\end{equation}}
\newcommand{\bea}{\begin{eqnarray}}
\newcommand{\eea}{\end{eqnarray}}
\newcommand{\bml}{\begin{mathletters}}
\newcommand{\eml}{\end{mathletters}}

\begin{document}

In the last few years progress has been achieved in studying string theory
in the presence of instabilities represented by tachyon fields. The problem is
particularly well understood when the tachyons appear in the open string 
spectrum \cite{Sen}.
In this case gravity can be ignored at weak string coupling and the problem
can be studied by a variety of means including string field theory
\cite{Kostelecky:1988ta,Sen:1999nx,Moeller:2000xv}, world-sheet renormalization group
(RG) flow \cite{Harvey:2000na} and  boundary string field theory \cite{Gerasimov:2000zp,
Kutasov:2000qp,Kutasov:2000aq,Ghoshal:2000gt}.

The study of closed string tachyon condensation is both more difficult and
more interesting. Open string tachyon condensation can be studied in a fixed,
often flat, spacetime background. In contrast, closed string tachyons couple
to other closed string modes, including the graviton, and hence the spacetime
geometry itself will generically change in the decay process. 

In some cases, closed string tachyons may be localized on defects such as
NS5-branes or orbifolds \cite{Adams:2001sv}. The localization makes the problem
more tractable and tachyon condensation in this system has been studied using
RG flow \cite{Adams:2001sv,Harvey:2001wm,Vafa:2001ra,Dabholkar:2001wn,
Gutperle:2002ki}, D-brane techniques
\cite{Adams:2001sv} and duality arguments \cite{Dabholkar:2001gz}. 
What has not been done so far is to study this process directly
in the supergravity r\'egime by finding time-dependent solutions which 
interpolate between one
conical spacetime at early times and another one at late times. 
That is what we shall do in this note.

The problem originally considered in \cite{Adams:2001sv} involves string
propagation on the orbifold spacetime
\be
R^{7,1} \times R^2/Z_n
\ee
That is, on eight-dimensional Minkowski space times a two-dimensional cone.
To be concrete we will consider IIA string theory on such a space.
String propagation on such an orbifold is exactly soluble and one finds
tachyonic modes in the twisted sector of the orbifold. These correspond to
instabilities which are localized at the fixed point of the orbifold, that
is at the tip of the cone. Any perturbation of the system which excites the
tachyon field should lead to a time-dependent solution of string theory
with the solution at late times approaching a new solution of string theory.
The full time dependent problem is currently too difficult to solve, so what
has been done instead is to replace time dependence with renormalization group
flow and to then study the corresponding RG equations. One result of these
studies is the claim that a $Z_n$ orbifold spacetime can decay to a $Z_{n'}$
spacetime with $n'<n$, that is to a cone with a smaller deficit angle (with
$n'=1$ being the stable case of flat space). 

Here we will take a different approach, studying the time-dependent equations
but in a low-energy limit. The mass of the tachyons in these theories is
typically string scale, so there is no consistent approximation in which we
keep the tachyon fields and ignore other string fields. However we can consider
a low-energy supergravity action with the tachyon fields simply acting as
source terms for the massless fields. The tachyon field will thus provide
initial conditions for the massless fields but we can then study the
time-evolution of the massless fields with these initial conditions. The
relevant massless fields are the graviton and dilaton. Thus we are led to
study the time-dependent equations for gravity coupled to a dilaton field.

To get a feel for the problem, we start by deriving a spacetime which consists of 
an expanding shell dividing two regions of differing conical deficits. Any two
spacetimes can be glued together across a wall, provided the 
induced metric on the wall is identical in each domain. The energy-momentum 
of the wall is then given by the jump in extrinsic curvature via the Israel 
conditions. In general a wall corresponds to a source, which will require 
some sort of matter; whether or not this can be achieved with the
fields at hand will be addressed presently.

Thus consider the conical spacetime
\be
ds^2 = dt^2 - dr^2 - dy_i^2 - r^2 (1-\Delta)^2 d\theta^2
\ee
Here the $y_i$ are coordinates on $R^7$ while $(r,\theta)$ are polar
coordinates for the two-dimensional cone. 
The angular variable, $\theta$, runs from $0$ to $2\pi$ on each side of
the wall for agreement of the internal wall metric, so if the deficit angle
differs, then clearly the value of $r$ on each side of the wall 
has to be different in order that $g_{\theta\theta}$ be the same. 
Note that this means $r_+ \propto r_-$.

We suppose that the wall has trajectory 
\be
x^\mu_\pm (\tau) = \left ( t_\pm(\tau), r_\pm(\tau) \right )
\ee
on each side, where ${\dot t}^2_\pm - {\dot r}^2_\pm = 1$. 
This has unit normal
\be
n_{\mu\pm} = \left ( {\dot r}_\pm, -{\dot t}_\pm \right )
\ee
which gives for the extrinsic curvature
\be 
K_{\tau\tau} = {\ddot r}/{\dot t} \qquad , \qquad K_{\theta\theta} 
= -{\dot t} r(1-\Delta)^2
\ee
The energy-momentum of the wall is given by Israel's equations, $T_{ab} 
= \delta K_{ab} - \delta K h_{ab}$, and reads:
\bea
T_y^y &=& -{{\ddot r}_+\over {\dot t}_+} + {{\ddot r}_-\over {\dot t}_-} 
-{{\dot t}_+\over r_+} + {{\dot t}_-\over r_-}\\
T_\tau^\tau &=& -{{\dot t}_+\over r_+} + {{\dot t}_-\over r_-} = 
T^y_y - T^\theta_\theta
\eea

So, for example, if $r = vt$, ${\dot t} = \gamma = 1/\sqrt{1-v^2}$, 
and ${\dot r} = v\gamma$.
The energy -momentum of this boundary is then
\bea
T_\tau^\tau =T^y_y &=& -{\gamma_+\over r_+} + {\gamma_-\over r_-}
= \left ( \gamma_-(1-\Delta_-) - \gamma_+(1-\Delta_+) \right ) 
[(1-\Delta)r]^{-1}\\ T_\theta^\theta &=& 0 
\eea
If we assume that the velocities on each 
side are the same (which is probably not necessary) then we have
\be
T_\tau^\tau =T^y_y  = {\gamma\over(1-\Delta r)} \left 
( \Delta_+ - \Delta_- \right )
\ee
Assuming that the exterior conical deficit exceeds the interior one, 
then this corresponds to a positive tension brane, but smeared over 
the $\theta$-direction, as it has no $\theta-\theta$ energy momentum.

This corresponds to an expanding region of lower conical deficit 
spacetime separated from the original spacetime by an expanding 
shell - a source. Whether or not this can be translated into a 
genuine solution to low-energy string theory necessitates solving the gravity 
equations of motion with dilaton source terms. We now show that this is 
possible in a certain limit of string theory.

Instead of solving the gravity problem in the low-energy limit of
ten-dimensional string theory, that is gravity 
with a dilaton, we consider an M-theory approach using eleven-dimensional gravity 
which renders the problem purely geometrical.
This approach is valid in the
strong coupling limit of ten-dimensional IIA string theory.

We look for a solution involving only the metric in eleven dimensions, \ie\ the 
three-form potential $C_{ABC} \equiv0$. We will also set the Kaluza-Klein (KK)
gauge field to zero, consistent with the RG equations, which can be consistently
truncated to those for only the metric and dilaton \cite{Adams:2001sv}.

We take spacetime to be the product of a seven-dimensional Ricci flat manifold
with coordinates $y^i$, the M-theory circle with coordinate $\psi$ and radius
determined by the dilaton $D$ in the usual way, and a $2+1$ dimensional spacetime
with coordinates $(t,r,\theta)$. 
For the latter we consider the most general time-dependent metric with
a single rotational degree of freedom, leading to the ansatz:
\be
ds^2 = B^2 (t,r) \left [ dt^2 - dr^2 \right]
-A^2 (t,r) d y_i^2 - C^2 (t,r) d\theta^2 - D^2 (t,r) d\psi^2
\label{genmet}
\ee

The Ricci curvature of  this metric is:
\bea
R^y_y &=& -{1\over B^2} \left [ {\partial_+\partial_- A\over A} 
+ 6{\partial_+A\partial_-A\over A^2} +  {1\over2} 
{\partial_+ A \over A}{\partial_-(CD)\over CD} + {1\over2} {\partial_-A\over A}
{\partial_+(CD)\over CD} \right ] \label{Ricciy}\\
R^\theta_\theta &=& -{1\over B^2} \left [ {\partial_+\partial_- C\over C}
+ {1\over2} {\partial_+ C \over C}{\partial_-(A^7D)\over A^7D} 
+ {1\over2} {\partial_-C\over C}
{\partial_+(A^7D)\over A^7D} \right ]
\label{Riccith} \\
R^\psi_\psi &=& -{1\over B^2} \left [{ \partial_+\partial_- D\over D}
+  {1\over2} {\partial_+ D \over D}{\partial_-(A^7C)\over A^7C} 
+ {1\over2} {\partial_-D\over D}
{\partial_+(A^7C)\over A^7C} \right ]
\label{Riccips} \\
R^t_t &=& {1\over B^2} \left [- 7{{\ddot A}\over A} - {{\ddot C}\over C}
- {{\ddot D}\over D} + {B'\over B}  {(A^7CD)'\over A^7CD} 
+{ {\dot B}\over B} {(A^7CD)^\cdot\over A^7CD} 
- \partial_+\partial_-(\ln B)\right ] \label{Riccit} \\
R^r_r &=& {1\over B^2} \left [ 7{A''\over A} + {C''\over C} + {D'' \over D}
- {B'\over B}  {(A^7CD)'\over A^7CD} 
-{ {\dot B}\over B} {(A^7CD)^\cdot\over A^7CD}  - \partial_+\partial_-
(\ln B)\right ]  \label{Riccir} \\
R_{rt} &=& -  {{\dot C}'\over C} - {{\dot D}'\over D} - 7{{\dot A}'\over A}
+ {{\dot B} \over B} {(A^7CD)'\over A^7CD} 
+ {B'\over B} {(A^7CD)^\cdot\over A^7CD} \label{Riccirt}
\eea
where $\partial_\pm = \partial/\partial x_\pm$, and $x_\pm = (t\pm r)/2$.

Although this form of the Ricci tensor does not look particularly illuminating, 
it is possible to rewrite the metric functions to produce instead a set of
free and interacting fields \cite{BCG} (one of which can be identified as the dilaton)
by defining:
\bea
\ln A &=& {\sigma\over7} \\
\ln B &=& \chi - \sigma - 2\phi/3\\
\ln D &=& 2\phi/3\\
C &=& \alpha e^{-\sigma} e^{-2\phi/3}
\eea

The vacuum equations now reduce to the much simpler form
\bea
\partial_+\partial_- \alpha &=& 0\\
\partial_+\partial_- \phi + {1\over2} \left ( 
\partial_+\phi {\partial_-\alpha\over\alpha}
+ \partial_-\phi{\partial_+\alpha\over\alpha} \right) &=& 0 \\
\partial_+\partial_- \sigma + {1\over2} 
\left ( \partial_+\sigma {\partial_-\alpha\over\alpha}
+ \partial_-\sigma{\partial_+\alpha\over\alpha} \right) &=& 0 \\
\partial_+\partial_- \chi + {4\over 9} \partial_+\phi \partial_-\phi
+ {4\over 7} \partial_+\sigma \partial_- \sigma 
+ {1\over 3} \partial_+\phi\partial_-\sigma
+ {1\over 3} \partial_+\sigma \partial_- \phi &=&0\\
{\partial_\pm^2 \alpha\over\alpha} + {8\over 9} 
\left ( \partial_\pm \phi \right )^2
+ {8\over7} \left ( \partial_\pm \sigma\right)^2 
+ {4\over3} \partial_\pm \phi\partial_\pm\sigma
-2\partial_\pm \chi {\partial_\pm \alpha\over\alpha} &=& 0
\eea
One way of viewing this redefinition is via a KK reduction
over $\{y_i,\theta,\psi\}$ to two dimensions. The free field, $\alpha$,
represents the overall volume of the internal space, with $\phi$
and $\sigma$ representing distortions, or relative volume breathing modes,
which couple to $\alpha$ via a friction term. $\phi$ is of course the
dilaton.

This now turns out to be totally prescriptive. Since there are 
no sources, we can simply make a coordinate choice to fix out 
the conformal coordinate freedom setting
\be
\alpha \equiv r
\ee
and then both $\phi$ and $\sigma$ satisfy cylindrically symmetric wave 
equations in 2+1 dimensions:
\be
{\ddot\phi} - \phi'' - {\phi'\over r} = 0 
= {\ddot\sigma} - \sigma'' - {\sigma'\over r}
\label{cyllap}
\ee
giving
\bea
\chi'  &=& r\left [ {4\over9} ({\dot\phi}^2 + \phi^{\prime2} ) + {4\over7}
({\dot\sigma}^2 + \sigma^{\prime2}) + {2\over3} (\sigma' \phi' 
+ {\dot\phi}~{\dot \sigma})\right] \label{chiprim} \\
{\dot\chi} &=& 2r\left [ {4\over9} {\dot\phi}\phi' +{4\over7}{\dot\sigma}\sigma'
+ {1\over3}( \sigma'{\dot\phi} + \phi'{\dot\sigma} ) \right]
\eea

These equations have some simple solutions.  If we set the dilaton and modulus of 
the extra seven dimensions to zero, then $\chi$ is obviously also a constant. 
Choosing $\chi=0$ gives flat spacetime. Choosing $\chi \ne 0$ results
in a conical spacetime, with deficit
\be
\Delta = 2\pi (1 - e^{-\chi_0})
\ee

We now want to find a solution which
is of this form for $r>t$, 
and has a form for $r<t$ which is asymptotic to a different
conical spacetime for $r\ll t$.
For simplicity we suppose that the extra $y$-dimensions 
in the string frame are inert \ie\ $A=D^{-1/2}$ 
or $\sigma = -7\phi/3$. This gives
\bea
\chi'  &=& 2r \left ( {\dot\phi}^2 + \phi^{\prime2} \right )  \\
{\dot\chi} &=& 4r  {\dot\phi}\phi' 
\eea

The general cylindrically symmetric
solution to the wave equation (\ref{cyllap}) with initial
data $\phi(0,r) = U(r)$, ${\dot \phi} (0,r) = V(r)$ is
\be
\label{dilsol}
\phi(\vec r,t) = \frac{1}{2\pi} \int_{|\vec r- \vec r'|\le t}
{V(r')d^2r'\over \sqrt{t^2-|\vec r - \vec r'|^2}}
+ \frac{1}{2\pi} {\partial\over\partial t}
\int_{|\vec r- \vec r'| \le t}
{U(r')d^2 r'\over \sqrt{t^2-|\vec r - \vec r'|^2}}
\ee

This shows that for a general dilaton wave, while the wavefront can
be largely clustered around the future lightcone, there will nonetheless
be a tail inside the future light cone, slowly settling down to a
vacuum solution. This is shown in figure \ref{fig:dpulse} where,
for illustrative purposes, we take $U(r)=0$ and $V(r)=e^{-r^2}$. 
\FIGURE{
\includegraphics[height=10cm]{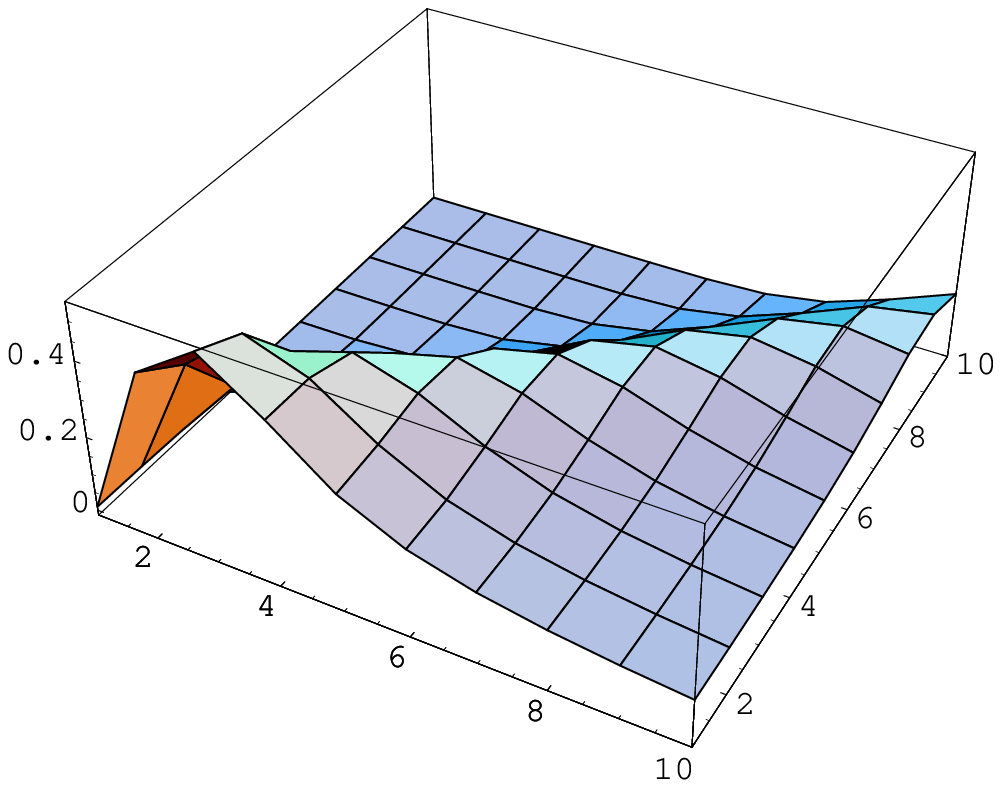}
\caption{Dilaton pulse as a function of $(r,t)$ for Gaussian initial
conditions}
\label{fig:dpulse}
}
Furthermore, if $U(r)$ and $V(r)$ are localized near the
origin, then (\ref{dilsol}) shows that at late times 
the fluctuations of the dilaton field disperse to infinity with the
the dilaton decaying back to its original value as $1/t$. This is
shown in figure \ref{fig:dilaton} for the same $U$ and $V$ as 
figure \ref{fig:dpulse}.
\FIGURE{
\includegraphics[height=8cm]{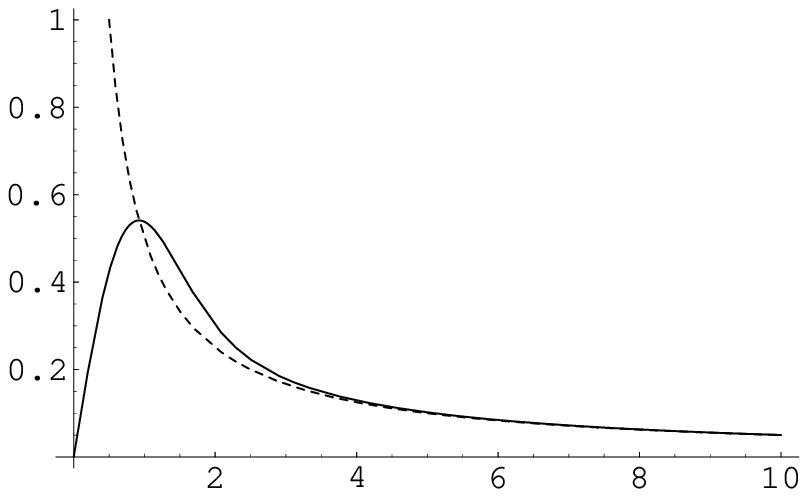}
\caption{The dilaton at the origin as a function of time. The
dotted line represents the $1/t$ asymptotic falloff for late times.}
\label{fig:dilaton}
}
Thus the decay process of the conical spacetime does not change the value
of the string coupling constant and hence can be studied consistently
at strong coupling.

Finally, a general result relating the two deficit angles is
obtained by integrating the $\chi'$ 
equation (\ref{chiprim}). Integrating out from $r=0$ we find that 
the right hand side is positive definite, independently
of the precise form of $\phi$ and $\sigma$. Thus $\chi$ will always be bigger 
at larger $r$ than at $r=0$. In other words, geometry mandates the
decay of larger deficits into smaller deficits as has been found in
other analyses of this problem \cite{Adams:2001sv,Harvey:2001wm}.

It would be interesting to interpret our results in terms of the change in 
Bondi energy. For a static conical spacetime the Bondi energy is proportional 
to the deficit angle. Our result on the decrease of the deficit angle under 
time evolution is suggestive of the decrease of Bondi energy with time in 
four-dimensional Einstein gravity \cite{Wald}.
In our case the outgoing energy would be carried away by the dilaton pulse 
rather than by gravitational radiation. However to our knowledge such an 
analysis has not been done in $(2+1)$-dimensional dilaton gravity, and indeed 
the usual analysis of asymptotic flatness is quite different
in odd spacetime dimensions \cite{Hollands:2003ie}. It may be
that the concept of C-energy \cite{Thorne} is the most appropriate 
tool for analysis of this system.

To summarize: we have studied the decay of conical spacetimes 
directly as time-dependent
solutions to gravity. Our results, which apply at strong string coupling,
are consistent with previous results based on RG flow and D-brane probe
analysis and show that these results can in part be understood purely in
gravitational terms. It would be interesting to extend this analysis to
higher dimensions, in particular to the orbifolds $C^2/Z_{n(p)}$ which
were studied in \cite{Adams:2001sv,Harvey:2001wm}. As might be expected from
the subtleties in the world-sheet analysis, this case is also much more
complicated from the space-time point of view. 

\acknowledgments
We would like to acknowledge the Aspen Center for Physics where 
most of this work was done.  We would also like to thank Patrick Dorey, 
Greg Moore, and John Stewart for helpful
discussions. RG was supported by the Royal Society, and JAH by NSF grant 
PHY-0204608.

\end{document}